# Spin Transport in Antiferromagnetic Insulators Mediated by Magnetic Correlations

Hailong Wang[†], Chunhui Du[†], P. Chris Hammel[*] and Fengyuan Yang[*]

Department of Physics, The Ohio State University, Columbus, OH, 43210, USA

[†]These authors made equal contributions to this work

[*]Emails: hammel@physics.osu.edu; fyyang@physics.osu.edu

We report a systematic study of spin transport in antiferromagnetic (AF) insulators having a wide range of ordering temperatures. Spin current is dynamically injected from $Y_3Fe_5O_{12}$ (YIG) into various AF insulators in Pt/insulator/YIG trilayers. Robust, long-distance spin transport in the AF insulators is observed, which shows strong correlation with the AF ordering temperatures. We find a striking linear relationship between the spin decay length in the AFs and the damping enhancement in YIG, suggesting the critical role of magnetic correlations in the AF insulators as well as at the AF/YIG interfaces for spin transport in magnetic insulators.

PACS: 75.50.Ee, 75.70.Cn, 76.50.+g, 81.15.Cd

Spin currents carried by mobile charges in metallic and semiconducting ferromagnetic (FM) and nonmagnetic (NM) materials have been the central focus of spintronics for the past two decades [1]. However, spin transport in AF insulators has been essentially unexplored due to the difficulty in generating magnetic excitations in these insulators. Ferromagnetic resonance (FMR) and thermally driven spin pumping [2-15] have attracted intense interest in magnon-mediated spin currents, which can propagate in both conducting and insulating FMs and AFs. We recently reported observation of highly efficient spin transport in AF insulator NiO with long spin decay length [16]. In this letter, we probe the mechanisms responsible for spin transport in AF insulators by investigating several series of Pt/insulator/YIG trilayers; this study is enabled by the large inverse spin Hall effect (ISHE) signals in our YIG-based structures [9-15].

Epitaxial YIG (epi-YIG) films are grown on (111)-oriented $Gd_3Ga_5O_{12}$ (GGG) substrates by sputtering [9-17]. X-ray diffraction and atomic force microscopy measurements reveal high crystalline quality and smooth surfaces of the YIG films [18]. Figure 1(a) shows an in-plane magnetic hysteresis loop for a 20-nm YIG film which exhibits a small coercivity ($H_c$) of 0.40 Oe and sharp magnetic reversal, indicating high magnetic uniformity. Figure 1(b) presents a FMR derivative absorption spectrum for a 20-nm YIG film taken in a cavity at radio-frequency (rf) $f = $ 9.65 GHz and microwave power $P_{rf} = 0.2$ mW with an in-plane magnetic field, which gives a narrow linewidth ($\Delta H$) of 7.7 Oe. All of these measurements are carried out at room temperature.

In order to probe spin transport in insulators of various magnetic structures, we select six materials, including: 1) amorphous $SrTiO_3$, a diamagnet, 2) epitaxial $Gd_3Ga_5O_{12}$, a paramagnet with a large magnetic susceptibility ($\chi$), and four antiferromagnets, 3) $Cr_2O_3$ [19], 4) amorphous YIG (*a*-YIG) [20], 5) amorphous $NiFe_2O_4$ (*a*-NFO) [21], and 6) NiO [19]. All insulator layers are deposited by off-axis sputtering. Lattice matched, strain-free $Gd_3Ga_5O_{12}$ films are epitaxially grown on YIG at high temperature; the remaining five insulators are grown at room temperature to avoid straining the epi-YIG films which can significantly alter the magnetic resonance in YIG. Electrical transport measurements confirm the highly insulating nature of all these films. Figures 1(a) and 1(b) indicate that the *a*-YIG film has negligible magnetization and FMR absorption (*a*-NFO films exhibit similar behavior). Thus, the six insulators include a diamagnet, a paramagnet, and four AFs with a wide range of ordering temperatures, allowing us to probe magnetic excitations and spin propagation in insulators both above and below the AF ordering temperatures, hence illuminating the roles of both static and dynamic magnetic correlations.

Bulk $Cr_2O_3$ and NiO have Néel temperatures $T_N$ = 318 and 525 K, respectively [19]. Both YIG and $NiFe_2O_4$ are ferrimagnets when in crystalline form; however, amorphous YIG and $NiFe_2O_4$ become AFs due to the lack of crystalline ordering required for ferrimagnetism [20, 21]. The temperature ($T$) dependence of exchange bias in FM/AF bilayers allows direct measurement of the blocking temperature, $T_b$, of the AFs. Our YIG(20 nm)/NiO(20 nm) bilayer exhibits a clear exchange bias field, $H_E$ = 13.5 Oe and an enhanced coercivity $H_c$ = 19.2 Oe ($H_c$ = 0.40 Oe for a single YIG film), demonstrating exchange coupling between YIG and NiO [18]. However, the very large paramagnetic background of GGG substrates prohibits the measurement of exchange bias at low temperatures needed for $Cr_2O_3$, *a*-YIG, and *a*-NFO. To determine $T_b$ for each AF studied here, we use $Ni_{81}Fe_{19}$ (Py) as the FM and measure exchange bias in Py(5 nm)/AF(20 nm) bilayers grown on Si. Figure 2(a) shows the hysteresis loops of four Py/AF bilayers at $T$ = 5 K after field cooling from above $T_b$. All four samples exhibit substantial exchange bias: $H_E$ = 646, 1403, 568, and 97 Oe for Py/NiO, Py/*a*-NFO, Py/*a*-YIG, and Py/$Cr_2O_3$, respectively. Figures 2(b) and 2(c) show the temperature dependencies of $H_E$ for the four bilayers, from which we determine $T_b$ = 20, 45, 70, and 330 K for $Cr_2O_3$, $a$-YIG, $a$-NFO, and NiO, respectively.

Spin currents in insulators propagate via precessional spin wave modes, e.g., magnons in ordered FMs and AFs. However, it is challenging to excite AF magnons which, for example, requires THz frequency in NiO [22]. Furthermore, the AF ordering temperatures in thin films decrease at lower thicknesses and conventional magnons cannot be sustained above the ordering temperatures. Here, we leverage the established technique of FMR spin pumping in YIG-based structures to excite the AF insulators via exchange coupling to the precessing YIG magnetization and to probe spin transfer in these insulators. For each of the six insulators, we grow a series of Pt(5 nm)/insulator($t$)/epi-YIG(20 nm) trilayers with various insulator thicknesses $t$ on YIG films cut from the same YIG/GGG wafer to ensure consistency of the YIG quality. Since Pt is the only conductor in the trilayers, the voltage signals detected are exclusively from the ISHE ($V_{ISHE}$), which proportionally reflects the spin currents pumped into Pt across the insulators.

Room-temperature spin pumping measurements [18] are conducted on all trilayers (~1 mm wide and ~5 mm long) in an FMR cavity at $f$ = 9.65 GHz and $P_{rf}$ = 200 mW in an in-plane DC field ($H$), as illustrated in Fig. 3(a). The mV-level ISHE voltages provide a dynamic range of more than three orders of magnitude for detecting the decay of spin current across the insulators. The rates at which $V_{ISHE}$ decays with increasing insulator thickness ($t$) differ dramatically among the six spacers. A 0.5-nm $SrTiO_3$ [18] already suppresses $V_{ISHE}$ by a factor of 17 from the corresponding Pt/YIG bilayer [10]. As we change the insulator from $SrTiO_3$ → $Gd_3Ga_5O_{12}$ → $Cr_2O_3$ → $a$-YIG → $a$-NFO → NiO, the spin currents exhibit substantially increasing propagation lengths. Figure 3(b) summarizes the $t$-dependencies of the normalized peak $V_{ISHE}$ at YIG resonance, $H_{res}$, for all six series. From the linear relationship in the semi-log plots, we extract the spin decay lengths $\lambda$ in the insulators by fitting to $V_{\text{ISHE}}(t)/V_{\text{ISHE}}(0) = e^{-t/\lambda}$, which gives $\lambda$ = 0.18, 0.69, 1.6, 3.9, 6.3, and 9.8 nm for SrTiO$_3$, Gd$_3$Ga$_5$O$_{12}$, Cr$_2$O$_3$, a-YIG, a-NFO, and NiO, respectively (Table I). More surprisingly, Fig. 3(c) shows that $V_{\text{ISHE}}$ initially increases by a factor of 2.1 and 1.6 when a 1- or 2-nm NiO and a-NFO, respectively, is inserted between YIG and Pt (the point for $t$ = 0 is excluded from the exponential fit for NiO and NFO). This dramatic variation in the spin current propagation length-scale most likely arises from different magnetic characteristics of the six insulators.

For dynamically generated spin current to transmit across insulating spacers beyond the tunneling range (~1 nm), magnetic excitations in the insulators are expected to play a major role. Except SrTiO$_3$, all other five insulators have strong magnetic character, including paramagnetic Gd$_3$Ga$_5$O$_{12}$ and four AFs with various ordering temperatures. For the same AF material, $T_b$ can vary significantly depending on the film thickness [23]. Among the four AFs, NiO is the most robust AF with $T_b$ = 330 K for our 20-nm NiO film, while for very thin NiO layers (<5 nm), $T_b$ is expected to be well below 300 K [24]. For a-NFO, a-YIG and Cr$_2$O$_3$, the AF ordering temperatures are well below room temperature (Table I). It is interesting to note that Gd$_3$Ga$_5$O$_{12}$ also exhibits magnetic order at very low temperatures [25]. Thus, magnetic correlations amongst thermally fluctuating AF moments are critically important for the observed spin transport in insulators.

These results suggest that, at resonance, the precessing YIG magnetization generates magnetic excitations in the adjacent insulator (either with static AF ordering or fluctuating correlated moments) via interfacial exchange coupling, which in turn enhances magnetic damping of the YIG. We measure the Gilbert damping constant $\alpha$ [26] from the frequency dependencies of FMR linewidths $\Delta H$ for six insulator(20 nm)/YIG(20 nm) bilayers and a single epi-YIG film using a microstrip transmission line. Figure 4(a) shows the linear frequency dependence of $\Delta H$ given by $\Delta H = \Delta H_0 + \frac{4\pi\alpha f}{\sqrt{3}\gamma}$ for the seven samples, where $\Delta H_0$ is the y-intercept and $\gamma$ is the gyromagnetic ratio. From the slopes of least-squares fits, we obtain $\alpha = 8.1 \times 10^{-4}$, $8.6 \times 10^{-4}$, $11 \times 10^{-4}$, $12 \times 10^{-4}$, $14 \times 10^{-4}$, $17 \times 10^{-4}$, $26 \times 10^{-4}$, and $36 \times 10^{-4}$ for the bare YIG, SrTiO$_3$/YIG, Gd$_3$Ga$_5$O$_{12}$/YIG, Cr$_2$O$_3$/YIG, a-YIG/YIG, a-NFO/YIG, NiO/YIG, and Pt/YIG, respectively (Table I). The diamagnetic SrTiO$_3$ does not enhance the damping of YIG within experimental uncertainty while its spin current decays over an atomic length scale ($\lambda = 0.18$ nm) due to quantum tunneling [10]. The large paramagnetic moments in Gd$_3$Ga$_5$O$_{12}$ can absorb angular momentum via exchange coupling to YIG and conduct spin current, resulting in a longer $\lambda = 0.69$ nm. The four AFs show much longer spin decay lengths together with enhanced damping of YIG due to strong magnetic correlations [23]. NiO more than triples the damping of YIG and its spin decay length is almost 10 nm, while clear spin current is detected over a NiO thickness of 100 nm.

The AF resonance frequency of NiO is about 1 THz [22] which is much higher than the 9.65 GHz used in our FMR excitation of YIG. Despite the difference in the dispersion relations of YIG and the AFs, our result clearly demonstrates highly efficient spin transport across the AFs. Considering that strong AF spin correlations have been observed well above $T_N$ for NiO [27], we believe the excitations responsible for spin transport in AFs must be magnons in ordered AFs and AF fluctuations in insulators with low blocking temperatures. In either case, the strongly correlated AF spins are excited via exchange coupling to the precessing YIG magnetization (either the net or staggered ferrimagnetic moments) at the AF/YIG interface and transfer the spin current across the insulator to the interface with Pt, where it is converted to a spin-polarized electron current in Pt. This is analogous to the predicted magnon currents in FM insulators [28].

The independently measured spin decay length $\lambda$ and damping enhancement $\Delta\alpha$ both increase monotonically following $SrTiO_3 \to Gd_3Ga_5O_{12} \to Cr_2O_3 \to a\text{-YIG} \to a\text{-NFO} \to NiO$ (Table I). Figure 4(b) further shows that $\lambda$ and $\Delta\alpha$ exhibit a nearly perfect linear relationship for all insulators excluding $SrTiO_3$. The excellent linear relationship between $\lambda$ and $\Delta\alpha$ of five significantly different insulators indicates that spin transfer across the YIG/AF interfaces (measured by $\Delta\alpha$) and spin propagation inside the AF insulators (characterized by $\lambda$) are tightly related. The exchange coupling between YIG magnetization and AF spins at the interfaces and the exchange interaction between adjacent AF spins within the AFs play a dominant role in spin transport in insulators.

Lastly, the strength of magnetic correlations depends on the AF ordering temperatures. Our results shown in Figs. 2 to 4 indicate that the correlation strength increases following the order $SrTiO_3$ (diamagnet) $\to Gd_3Ga_5O_{12}$ (paramagnet) $\to Cr_2O_3$ (AF, $T_b = 20$ K)$\to a$-YIG (AF, $T_b = 45$ K) $\to a$-NFO (AF, $T_b = 70$ K) $\to$ NiO (AF, $T_b = 330$ K). As magnetic correlation increases, exchange interaction becomes stronger, which, 1) facilitates the propagation of spin currents carried by magnetic excitations in the insulators, and 2) enhances the magnetic damping of the underlying YIG films. The surprising enhancement of ISHE signals for the trilayers with 1- or 2-nm NiO and $a$-NFO [Fig. 3(c)] indicates that the Pt/NiO/YIG and Pt/$a$-NFO/YIG trilayer structures are highly efficient in spin transfer, while the underlying mechanism remains to be understood.

In summary, we observe clear spin currents in AF insulators mediated by AF magnetic correlations be they static or fluctuating. This result brings a large family of insulators, in particular, AF insulators, into the exploration of spintronic applications utilizing pure spin currents.

This work was primarily supported by the U.S. Department of Energy (DOE), Office of Science, Basic Energy Sciences, under Grants No. DEFG02-03ER46054 (FMR and spin pumping characterization) and No. DE-SC0001304 (sample synthesis and magnetic characterization). This work was supported in part by the Center for Emergent Materials, an NSF-funded MRSEC, under Grant No. DMR-1420451 (structural characterization). Partial support was provided by Lake Shore Cryogenics, Inc., and the NanoSystems Laboratory at the Ohio State University.

**Table I**. Type of magnetism, blocking temperatures (for AFs only), and spin decay length ($\lambda$) for the six insulators as well as the Gilbert damping constant ($\alpha$) for the six insulators (20-nm) on YIG. The parameters for a single epitaxial YIG film are also included for comparison.

| Layer | Magnetism | $T_b$ (K) | $\lambda$ (nm) | $\alpha$ |
|---|---|---|---|---|
| epi-YIG | ferrimagnet | | | $(8.1 \pm 0.6) \times 10^{-4}$ |
| SrTiO$_3$ | diamagnet | | $0.18 \pm 0.01$ | $(8.6 \pm 1.0) \times 10^{-4}$ |
| Gd$_3$Ga$_5$O$_{12}$ | paramagnet | | $0.69 \pm 0.02$ | $(11 \pm 1) \times 10^{-4}$ |
| Cr$_2$O$_3$ | antiferromagnet | 20 | $1.6 \pm 0.1$ | $(12 \pm 1) \times 10^{-4}$ |
| *a*-YIG | antiferromagnet | 45 | $3.9 \pm 0.2$ | $(14 \pm 1) \times 10^{-4}$ |
| *a*-NFO | antiferromagnet | 70 | $6.3 \pm 0.3$ | $(17 \pm 2) \times 10^{-4}$ |
| NiO | antiferromagnet | 330 | $9.8 \pm 0.8$ | $(26 \pm 3) \times 10^{-4}$ |

**Figure captions:**

**Figure 1**. (a) Room temperature in-plane magnetic hysteresis loops of a 20-nm epitaxial YIG film (blue) and a 20-nm amorphous YIG film (red) grown on GGG, where the paramagnetic background is from the GGG substrate. Inset: low-field hysteresis loop of the epitaxial YIG film showing a coercivity of 0.40 Oe. (b) FMR derivative absorption spectra of an epitaxial (blue) and an amorphous (red) 20-nm YIG film on GGG.

**Figure 2.** (a) Magnetic hysteresis loops of four Py(5 nm)/AF(20 nm) bilayers at 5 K after field cooling, all demonstrating clear exchange bias. Temperature dependence of $H_E$ for the four Py(5 nm)/AF(20 nm) bilayers in (b) linear and (c) log y-scale give the AF blocking temperature $T_b$ = 20, 45, 70, and 330 K for $Cr_2O_3$, $a$-YIG, $a$-NFO, and NiO, respectively.

**Figure 3.** (a) Schematic of the ISHE measurement on various Pt/Insulator/YIG structures. (b) Semi-log plots of $V_{ISHE}$ as a function of the insulator thickness for the six series normalized to the values for the corresponding Pt/YIG bilayers, where the straight lines are exponential fits to each series, from which the spin decay lengths $\lambda$ are determined. (c) Details of behavior shown in (b) for insulators below 10 nm.

**Figure 4.** (a) Frequency dependencies of FMR linewidths of a bare epitaxial YIG film, $SrTiO_3$(20 nm)/YIG, $Gd_3Ga_5O_{12}$(20 nm)/YIG, $Cr_2O_3$(20 nm)/YIG, $a$-YIG/(20 nm)/YIG, $a$-NFO(20 nm)/YIG, and NiO(20 nm)/YIG bilayers. (b) Excellent linear correlation between spin decay length $\lambda$ and Gilbert damping enhancement $\Delta\alpha = \alpha_{\text{Insulator/YIG}} - \alpha_{\text{YIG}}$. The line is a least-squares linear fit to all data points excluding $SrTiO_3$.

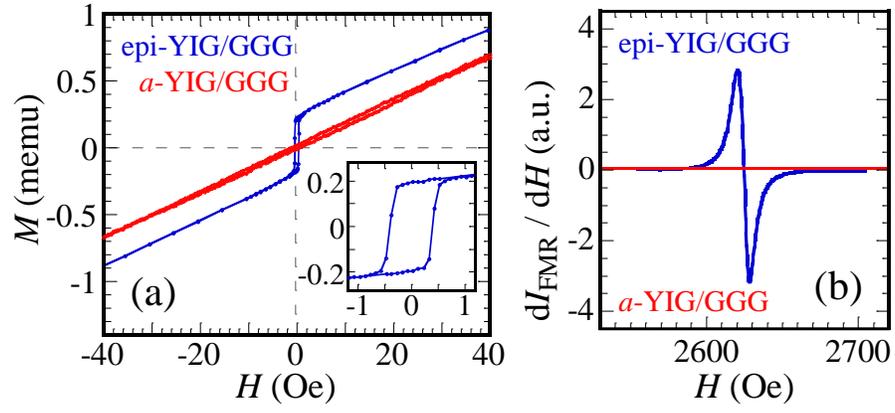

**Figure 1**

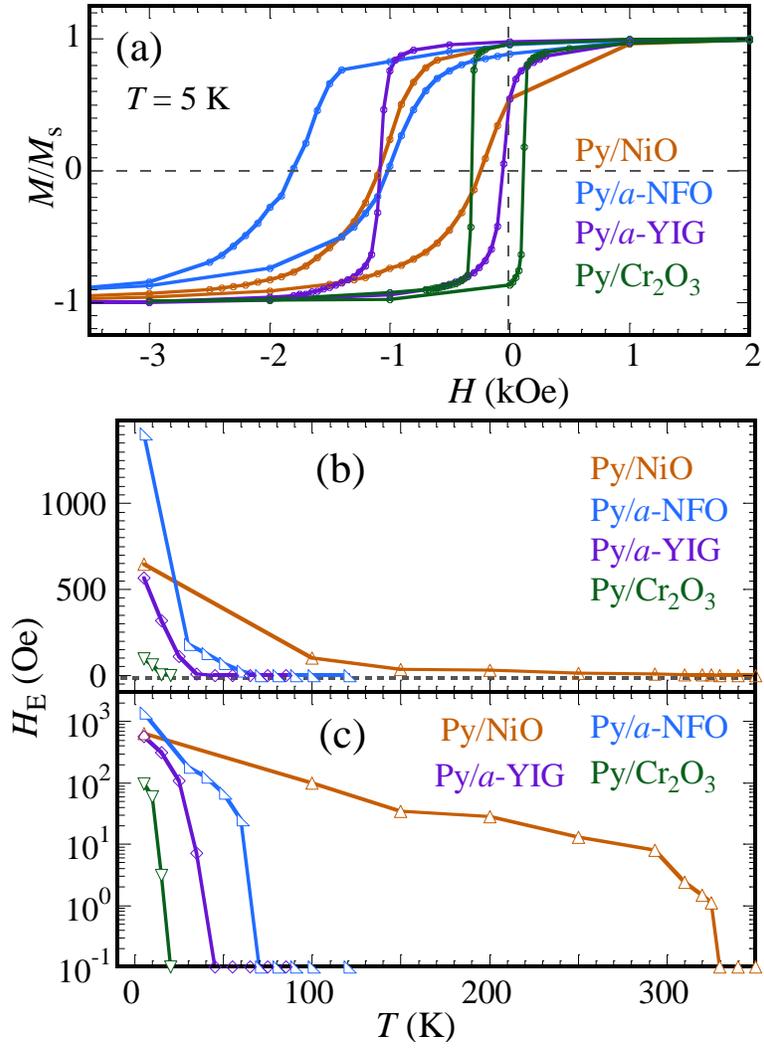

**Figure 2**

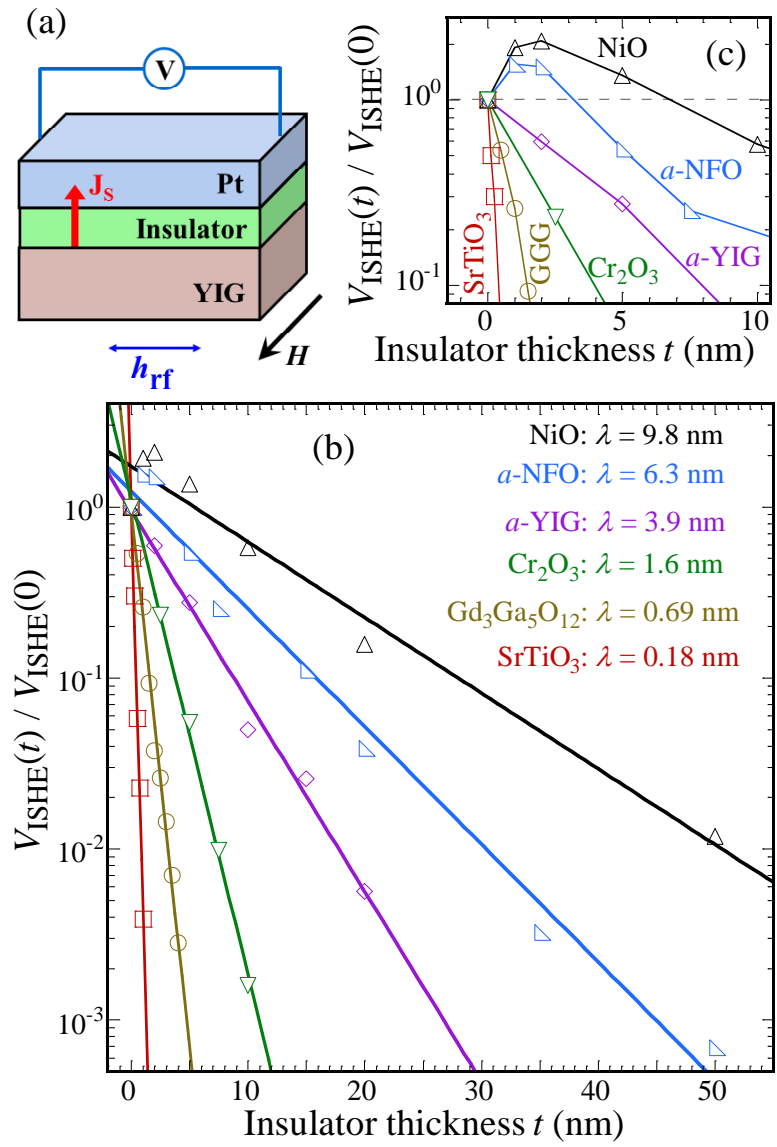

**Figure 3**

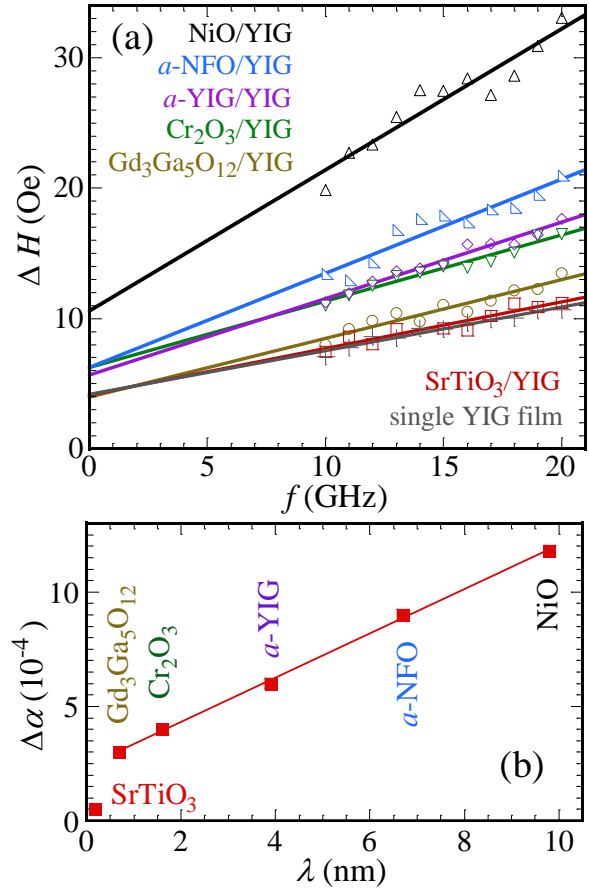

**Figure 4**